\documentclass[12pt]{article}
\usepackage{latexsym}
\usepackage{epsfig}

\def\bg#1{\mbox{\boldmath$#1$}}

\newcommand{\del}{\partial}
\newcommand{\beq}{\begin{eqnarray}}
\newcommand{\eeq}{\end{eqnarray}}
\newcommand{\be}{\begin{eqnarray*}}
\newcommand{\ee}{\end{eqnarray*}}
\newcommand{\bk}{{\bf k}}
\newcommand{\bp}{{\bf p}}

\newcommand{\bx}{{\bf x}}

\newcommand{\ra}{\rightarrow}

\newcommand{\ve}{\varepsilon}

\newcommand{\wh}[1]{\widehat{#1}}

\newcommand{\om}{{\omega}}

\begin{document}

\centerline{\Large\bf{Electromagnetic energy-momentum tensors in media}}
\bigskip
\centerline{ Finn Ravndal\footnote{On sabbatical leave of absence from  Institute of Physics, University of Oslo, N-0316 Oslo, Norway.}}
\bigskip
\centerline{\it Department of Physics, University of Miami, Coral Gables, FL 33124.}

\begin{abstract}
\small{It is pointed out that the previous energy-momentum tensors of Minkowski and Abraham for the electromagnetic field in continuous media are based on a covariant formulation which does not reflect a symmetry inherent to the system. Instead, taking into account  the intrinsic invariance under Lorentz transformations involving the reduced speed of light in such a medium, a compact and fully consistent theory can be formulated without the old problems.}
\end{abstract}

\section{Introduction}

After close to a hundred years, there are still two possible energy-momentum tensors  for the electromagnetic field in continuous media being discussed in the literature\cite{previous}. The first was derived by Minkowski\cite{Minkowski} and the second proposed shortly afterwards by Abaham\cite{Abraham}. Both of them have problems, in particular when they are used in a quantum context. In most textbooks these difficulties are only hinted upon. The book by Panofsky and Phillips\cite{PP} endorses the Abraham energy-momentum tensor. In the latest edition of the book by Jackson\cite{Jackson}  one is of the same view although one is open for the possibility that there might be an additional co-traveling momentum from the mechanical momentum of the electrons which could add up to the Minkowski electromagnetic momentum density. The same support for the Abraham version is also presented in the book by Landau and Lifshitz\cite{LL}, based on the usual requirement of relativistic invariance. 

A clear presentation of the original ideas behind the  theories can be found in the book by M\o ller\cite{Moller}. Both of them are covariantly formulated, based on Lorentz transformations in the vacuum. But this apparent Lorentz invariance is not an intrinsic symmetry of the medium where light moves with a reduced velocity. Taking into account this physical fact, an effective theory has recently been proposed where these problems are avoided\cite{EFT}. The free theory can be quantized by standard methods and extended with higher-order interaction terms to also describe dispersive and Kerr effects. It thus becomes a full-fledged effective field theory for electromagnetic phenomena in media and relates many classical and quantum effects in a systematic way. 

When light moves from the vacuum into an isotropic and transparent medium, it is in general refracted. Its frequency $\nu$  remains the same, but the wavelength $\lambda$ is reduced by the refractive index $n>1$. The corresponding phase velocity $\nu\lambda$ is therefore lowered to $1/n$ when we set the light velocity in vacuum to $c=1$. But it should not be forgotten that this is an effective description valid on large scales where the discrete atoms in the material can be replaced by a continuous medium. On the atomic scale light is moving with the vacuum velocity $c=1$ between interactions with electrons around the atoms. These will scatter the light in such a way that in the forward direction the scattered waves add up to a plane wave. However, it is delayed by  a phase shift of $\pi/2$ relative to the incoming wave as for instance explained by Feynman\cite{RPF}. The interference between these two waves will then effectively slow down the propagating wave.  As a result of these microscopic processes, the resulting wave is a highly complex object. In spite of that, experience shows that we can describe it by local electromagnetic fields obeying the standard Maxwell's equations for continuous media. But these are now effective fields, incorporating complicated physics on very short scales.

For this reason one would think that we today have a satisfactory and consistent theory for electromagnetic phenomena in media. And to a large extent that is certainly true. But the energetics of these processes are still unclear, caused by the uncertainty about the energy-momentum tensor. The Minkowski tensor is not symmetric and therefore has problems with angular-momentum conservation\cite{Minkowski}.   For this reason Abraham proposed a symmetric tensor at the cost of having to introduce a new, electromagnetic volume force in the medium\cite{Abraham}.  Its existence has still not been verified experimentally. The new tensor is not conserved and results in a momentum density of the field smaller than the Minkowski value by a factor $n^2$. Since then several other energy-momentum tensors have been suggested.  Experimentally, the different proposals are most directly tested by their predictions for the radiation pressure\cite{pressure}. This seems to favor the original construction of Minkowski. The different theories and experiments were reviewed in detail some time ago by Brevik\cite{Brevik}. A more recent survey of the whole situation can be found in the more phenomenological approach by Garrison and Chiao\cite{Chiao}.

We will in the following sum up the essence of the Minkowski and Abraham theories and comment in more detail upon their properties. At the end, the main idea behind the recently proposed effective field theory will be discussed\cite{EFT}. It avoids these problems and can also be quantized in a straight-forward way. Its physical properties are close to those of the Minkowski theory which is consistent with most experiments.

\section{Electromagnetism in media}

Without any charges or currents present in the material,  the electric fields  ${\bf E}, {\bf D}$  and magnetic fields ${\bf B}, {\bf H}$ are in general governed by the  Maxwell equations
\beq
          {\bg\nabla}\times{\bf E} + {\del{\bf B}\over\del t} = 0, \hspace{10mm}      {\bg\nabla}\cdot{\bf B} = 0                     \label{Max-1}
\eeq
and
\beq
          {\bg\nabla}\times{\bf H} - {\del{\bf D}\over\del t} = 0,     \hspace{10mm}    {\bg\nabla}\cdot{\bf D} = 0                 \label{Max-2}
\eeq
The displacement field ${\bf D}$ describes the modification of the electric field ${\bf E}$ by the polarization of the atoms in the material, while ${\bf H}$ describes the similar modification of the magnetic field ${\bf B}$ due to magnetization of the atoms. 

These are in general local fields rapidly varying on the microscopic scale between the atoms. On this scale we expect ordinary Lorentz invariance to be valid. Indeed, it can explicitly verified by writing these four equations on a covariant form. For that purpose, let the fields ${\bf E}$  and ${\bf B}$ be the components of the standard antisymmetric tensor
\beq
           F^{\mu\nu} = \left( \begin{array}{c | c} 0 & - {\bf E} \\ \hline  {\bf E} & - B_{ij}\end{array}\right)          \label{F}
\eeq
when we write $B_{ij} = \ve_{ijk} B_k$. Then we can write the first two Maxwell equations (\ref{Max-1}) as
\beq
            \del_\lambda F_{\mu\nu} +  \del_\nu F_{\lambda\mu} +  \del_\mu F_{\nu\lambda} = 0         \label{max-1}
\eeq
where the covariant gradient operator has the components $\del_\mu = (\del_t, {\bg\nabla})$. Similarly, combining ${\bf D}$  and ${\bf H}$ into the corresponding antisymmetric tensor
\beq
           H^{\mu\nu} = \left( \begin{array}{c | c} 0 & - {\bf D} \\ \hline {\bf D} & - H_{ij}\end{array}\right)          \label{H}
\eeq
with $H_{ij} = \ve_{ijk} H_k$, the two last equations (\ref{Max-2}) become simplified to
\beq
           \del_\nu H^{\mu\nu} = 0                               \label{max-2}
\eeq 
Had we included a four-current $J^\mu = (\rho,{\bf J})$ coupled to these fields, it would have appeared on the right hand side of this equation.

In the rest frame of the medium the fields are related by ${\bf D} = \ve {\bf E}$ and ${\bf B} = \mu {\bf H}$ for isotropic matter. Both the electric permittivity $\ve$ and the magnetic permeability $\mu$ will be taken to be constants.  These constitutive relations now represent an effective description of the fields in the medium where microscopic variations are averaged out in some sense. 

These relations can also be formulated in an arbitrary frame as already shown by Minkowski\cite{Minkowski}.  Let the medium move with velocity ${\bf v}$ and therefore with corresponding four-velocity $V^\mu =\gamma(1,{\bf v})$.  Here $\gamma = (1 - v^2)^{-1/2}$ as follows from $V^\mu V_\mu =1$.  Defining $F^\mu = F^{\mu\nu}V_\nu$, one can then write
\be
               \mu H_{\mu\nu} = F_{\mu\nu} + (n^2-1)(F_\mu V_\nu - F_\nu V_\mu)            
\ee
where now $n^2 = \ve\mu$. Several years later Gordon\cite{Gordon} rewrote this on the more suggestive form
\beq
            \mu H_{\mu\nu}  = X^{\rho}_{\,\,\,\mu} X^{\sigma}_{\,\,\,\nu} F_{\rho\sigma}    \label{const}
\eeq
after introducing the tensor
\beq
               X_{\mu\nu}  =  \eta_{\mu\nu}  + (n^2 - 1)V_\mu V_\nu             \label{X}
\eeq
The last term $\propto (n^2-1)^2$ in (\ref{const}) will not contribute because of the antisymmetry of $F_{\rho\sigma}$. Here $\eta_{\mu\nu}$ is the Minkowski metric so that $X_{\mu\nu}$ plays the role of a generalized metric\cite{AM}.

In order to find an equation of motion of the field, one must combine the two covariant Maxwell equations (\ref{max-1}) and (\ref{max-2}). The first is satisfied as an identity by introducing a four-vector potential $A^\mu$ so that $F_{\mu\nu} = \del_\mu A_\nu - \del_\nu A_\mu$. Using then the constitutive relation (\ref{const}) in the second equation (\ref{max-2}), the corresponding wave equation follows\cite{BL}. It takes the simplest form when one chooses a gauge equivalent to the Lorenz gauge in vacuum, i.e.
\beq
                  X^{\mu\nu}\del_\mu A_\nu = \del_\mu A^\mu + (n^2-1)(V^\mu\del_\mu)(A^\nu V_\nu) = 0         \label{gauge}
\eeq
The wave equation then becomes
\beq
               X^{\rho\sigma}\del_\rho\del_\sigma A_\mu = [\del^2 + (n^2 -1)(V\cdot\del)^2] A_\mu= 0          \label{waveq}
\eeq
In the rest frame of the medium  where $V^\mu = (1, 0)$ it takes the simpler form
\beq
           \Big(n^2{\del^2\over\del t^2} - \nabla^2\Big)A_\mu = 0                       \label{rest-waveq}
\eeq
An electromagnetic disturbance will therefore move with the phase velocity $1/n$ as expected in this frame. The gauge condition (\ref{gauge}) in the rest frame similarly reduces to
\beq
             n^2{\del\Phi\over \del t}  + {\bg\nabla}\cdot{\bf A} = 0        \label{restgauge}
\eeq
when we write $A^\mu = (\Phi, {\bf A})$.

One can also derive the wave equation from a variational principle. The  Lagrangian follows from the requirement that the vacuum theory should result in the limit $n\ra 1$ when $H_{\mu\nu}  \ra F_{\mu\nu}$. It then becomes clear that the simplest choice
\beq
               {\cal L} = - {1\over 4} F_{\mu\nu}H^{\mu\nu}                \label{Lag.1}
\eeq
has the desired properties.  In terms of the electric and magnetic vector fields it becomes ${\cal L} =  ({\bf E} \cdot{\bf D} - {\bf B} \cdot{\bf H})/2$ when inserting the tensors (\ref{F}) and (\ref{H}). Using the constitutive relations in the rest frame of the medium, it is
\beq
            \mu {\cal L} =  {1\over 2} \big(n^2{\bf E}^2 - {\bf B}^2\big)                \label{Lag.3}
\eeq
In a general frame one must instead use the constitutive relations on the covariant form (\ref{const}). The Lagrangian is then a function of the 4-vector potential $A_\mu$, but will also depend explicitly on the four-velocity  $V_\mu$ of the medium. It is therefore not strictly Lorentz invariant, in spite of its covariant look in ({\ref{Lag.1}).

\section{The Minkowski tensor}

Most of the physical content of the field theory is contained in its energy-momentum tensor. In the present case it was first obtained by Minkowski\cite{Minkowski}. It can be derived directly from the four Maxwell equations (\ref{Max-1}) and (\ref{Max-2}) as shown in the book by M\o ller\cite{Moller}. One then finds the tensor
\beq
                 T^{\mu\nu}_M =  F^\mu_{\,\,\,\alpha}H^{\alpha\nu} + {1\over 4}\eta^{\mu\nu} F_{\alpha\beta}H^{\alpha\beta}      \label{T_M}
\eeq
Its components can be displayed in the matrix
\beq
           T^{\mu\nu}_M =  \left( \begin{array}{c | c} {\cal E} & {\bf N} \\ \hline  {\bf G} &T_{ij}\end{array}\right)        \label{T_Mmat}
\eeq
The component $T^{00}$ is the energy density 
\beq
          {\cal E} =  {1\over 2} \big({\bf E} \cdot{\bf D} + {\bf B} \cdot{\bf H}\big)          \label{energy}
\eeq
while $T^{0k}$ form the components of the Poynting vector ${\bf N} = {\bf E}\times{\bf H}$. The tensor is seen to be non-symmetric with the components $T^{k0}$ forming the similar vector ${\bf G} = {\bf D}\times{\bf B}$. Finally, the space-space components make up the tensor
\beq
                T_{ij} =  - (E_i D_j + B_i H_j) + {1\over 2}\delta_{ij}({\bf E} \cdot{\bf D} + {\bf B} \cdot{\bf H})     \label{max-stress}
\eeq
Except for the sign, this becomes the  symmetric Maxwell stress tensor when we use the constitutive relations in the rest frame of the medium.

It should be noted here that the energy-momentum tensor (\ref{T_M}) has been derived without the use of the constitutive relations. Thus the macroscopic description implied by these and which are necessary when we want to describe the system at large scales, have not been taken into account. For that reason one might think that the tensor describes the energy-momentum content of the theory at the microscopic level, before any averaging has taken place. This could also have to do with it not being symmetric.

Using the two Maxwell tensor equations (\ref{max-1}) and (\ref{max-2}) it is straight forward to show that the Minkowski energy-momentum tensor
(\ref{T_M}) is conserved on the second index, i.e. $\del_\nu T^{\mu\nu}_M = 0$. For $\mu = 0$ this implies
\beq
      {\del{\cal E}\over\del t} + {\bg\nabla}\cdot{\bf N} = 0        \label{energy-cons}
\eeq
which is the standard expression for energy conservation. Similarly, for $\mu$ in a spatial direction, it gives
\beq
        {\del{\bf G}\over\del t} + {\bg\nabla}\cdot{\bf T} = 0    \label{mom-cons}
\eeq
where the second term is the divergence of the tensor (\ref{max-stress}). This equation now represents momentum conservation when ${\bf G}={\bf D}\times{\bf B}$ is taken to be the momentum density of the electromagnetic field. Needless to say, this is also the standard momentum density resulting from a non-covariant derivation\cite{PP}\cite{Jackson}.

It is well known after Einstein's construction of the general theory of relativity\cite{Einstein} that a general Lorentz-invariant theory can be made invariant under arbitrary coordinate transformations by replacing the flat metric $\eta_{\mu\nu}$ with a general metric  $g_{\mu\nu}$, letting partial derivatives $\del_\mu$ be replaced by covariant derivatives $\nabla_\mu$ etc. Hilbert showed in this connection\cite{Hilbert} that the energy-momentum tensor for such a system with Lagrangian ${\cal L}$ can then always be obtained from
\beq
           T_{\mu\nu} =  2{\del{\cal L}\over\del g^{\mu\nu}} - g_{\mu\nu}{\cal L}           \label{T_H}
\eeq
To return to flat Minkowski spacetime, one then lets $g_{\mu\nu}\ra \eta_{\mu\nu}$. This approach will obviously give a symmetric result. And the coordinate invariance implies that it is conserved. If we now follow this procedure for the electromagnetic Lagrangian (\ref{Lag.1}), it is easy to show that it does not reproduce the Minkowski tensor, but instead the symmetric combination $T^{\mu\nu} = (T^{\mu\nu}_M + T^{\nu\mu}_M)/2$. But this is now neither conserved on the first nor the second index. We can understand this from the earlier observation that the Lagrangian (\ref{Lag.1}) is not Lorentz invariant, in spite of its appearance. It is just written as a covariant version of the fundamental, rest-frame Lagrangian (\ref{Lag.3}).

\section{The Abraham tensor}

One can split up the Minkowski tensor (\ref{T_Mmat}) in two terms,
\be
          T^{\mu\nu}_M =   T^{\mu\nu}_A +  (n^2 - 1)\left( \begin{array}{c | c} 0 & 0\\ \hline  {\bf N} & 0 \end{array}\right)   
\ee
where the first part
\beq
            T^{\mu\nu}_A =  \left( \begin{array}{c | c} {\cal E} & {\bf N} \\ \hline  {\bf N} &T_{ij}\end{array}\right)      \label{T_A}
\eeq
is the Abraham tensor\cite{Abraham}. It is symmetric by construction. From the above conservation law for the the Minkowski tensor on the second index, it follows that the Abraham tensor is not generally conserved. Instead, it  satisfies $\del_\nu T^{\mu\nu}_A +  K^\mu = 0$ where $K^\mu$ is a force density. In the same frame as above we have $K^\mu = (0,{\bf K})$ where
\beq
         {\bf K} = (n^2 - 1){\del\over\del t}({\bf E}\times{\bf H})           \label{A-force}
\eeq
is the Abraham force. The time component of this new conservation law ensures energy conservation on the standard form (\ref{energy-cons}). While the spatial components again ensure momentum conservation, the momentum density of the field is now seen to be ${\bf E}\times{\bf H}$ and therefore $n^2$ smaller than the above Minkowski density ${\bf D}\times{\bf B}$. According to Garrison and Chiao\cite{Chiao}  the Abraham formalism seems to be needed to explain a few experiments where the systems under investigation undergo acceleration.

\section{Effective field theory}

As mentioned in the first section, one can write the Lagrangian (\ref{Lag.3}) in a general frame characterized by its 4-velocity $V^\mu$ on a form which looks invariant under ordinary Lorentz transformations. However, this covariant formulation does not represent a real invariance of the theory. Instead, it has another, built-in invariance represented by the Lorentz group corresponding to the light speed $1/n$ in the medium. We call it the material Lorentz group. It is a symmetry of the electromagnetic Lagrangian. But it does not represent an invariance of the whole system since the medium itself is not invariant under the corresponding transformations. Vacuum electromagnetism is in this respect different since the vacuum is invariant under ordinary  Lorentz transformations.

In order to make this material symmetry manifest, introduce the corresponding covariant coordinate four-vector $x^\mu = (t/n, \bx)$ so that $\del_\mu = (n\del_t, {\bg\nabla})$.  For the same reason we form the new four-vector potential $A^\mu = (n\Phi, {\bf A})$. The electric and magnetic field vectors are still given by the antisymmetric tensor $F_{\mu\nu} = \del_\mu A_\nu - \del_\nu A_\mu$ which now has the components
\beq
           F^{\mu\nu} = \left( \begin{array}{c | c} 0 & -n {\bf E} \\ \hline  {n\bf E} & - B_{ij}\end{array}\right)          \label{new-F}
\eeq
instead of (\ref{F}).  In terms of this, the rest-frame Lagrangian (\ref{Lag.3}) takes the standard form $\mu{\cal L} = - (1/4)F_{\mu\nu}^2$. It is now explicitly invariant under material Lorentz transformations. The first set of field equations (\ref{max-1}) obviously remains unchanged while the second Maxwell equation (\ref{max-2}) is replaced by $\del_\mu F^{\mu\nu} = 0$ when we make use of the con\-stitutive equations.  Writing this out, it becomes $\del^2 A^\nu = \del^\nu(\del\cdot A)$ where now $\del^2 = n^2\del_t^2 - \nabla^2$. In the Lorenz gauge $\del_\mu A^\mu = 0$,  which is seen to be equal to (\ref{restgauge}), this gives the previous wave equation (\ref{rest-waveq}). 

One can now also find a satisfactory energy-momentum tensor from the Hilbert prescription (\ref{T_H}) appropriate to this new Lorentz group. It is
\beq
                \mu T^{\mu\nu} =  F^\mu_{\,\,\,\alpha}F^{\alpha\nu} + {1\over 4}\eta^{\mu\nu} F_{\alpha\beta}F^{\alpha\beta}      \label{T_eff}
\eeq
with components
\beq
           T^{\mu\nu} =  \left( \begin{array}{c | c} {\cal E} & n {\bf N} \\ \hline n {\bf N} &T_{ij}\end{array}\right)        \label{T_effmat}
\eeq
It is obviously symmetric, traceless and conserved on both indices, i.e. $\del_\mu T^{\mu\nu} = \del_\mu T^{\nu\mu} = 0$. In the time direction this gives energy conservation on the form (\ref{energy-cons}) while in the space directions it gives momentum conservation as in (\ref{mom-cons}). The momentum density of the field ${\bf G} = {\bf D}\times{\bf B}$ is therefore the same as in the Minkowski theory.  

In the rest system one can consider the Lagrangian (\ref{Lag.3}) describing an electromagnetic disturbance in the medium on an equal footing with the Lagrangians for similar sound or spin excitations in matter.  For a plane wave of the form $e^{i(\bk\cdot\bx - \om_\bk t)}$ the corresponding wave equation (\ref{rest-waveq}) gives $\om_\bk = |\bk|/n$. Such linear dispersion relations occur often  in the long-wave limit for excitations in  condensed-matter physics. The rest frame of such a system is a preferred frame. It is also in this frame it is most natural to quantize the excitations. 

For the electromagnetic field under consideration the quantization can be done by elementary methods. In the Coulomb gauge ${\bg\nabla}\cdot{\bf A} = 0$ we expand the vector field ${\bf A}(\bx,t)$  in plane waves satisfying periodic boundary conditions. Each Fourier mode ${\bf A}_\bk(t)$ is transverse to the wave vector $\bk$.  We thus have only two polarization directions. From the Lagrangian (\ref{Lag.3}) each such mode then obeys the dynamics of a harmonic oscillator with frequency $\om_\bk = |\bk|/n $.  Integrating the above energy density, we find the Hamiltonian for the quantized photon field on the standard form
$\wh{H} = \sum_{\bk\lambda} \hbar\om_\bk \big(\wh{a}_{\bk\lambda}^\dagger\wh{a}_{\bk\lambda}  +1/2 \big)$ when we set the permeability $\mu =1$ as for a dielectric. The last term is the zero-point energy. Here the operator $\wh{a}_{\bk\lambda}$ annihilates a photon with wave vector $\bk$ and polarization $\lambda$. Together with the adjoint creation operator $\wh{a}_{\bk\lambda}^\dagger$ they satisfy the canonical commutator relation $[\wh{a}_{\bk\lambda},\wh{a}_{\bk'\lambda'}^\dagger] = \delta_{\bk\bk'} \delta_{\lambda\lambda'}$. A photon with wave number $\bk$ thus has the energy $E = \hbar\om_\bk$. Similarly, from the total field momentum ${\bf P} = \int\! d^3x {\bf D}\times{\bf B}$ one obtains the momentum operator
$\wh{\bf P} = \sum_{\bk\lambda}\hbar\bk\wh{a}_{\bk\lambda}^\dagger\wh{a}_{\bk\lambda}$ after quantization.  As expected, the photon with wave vector $\bk$  has the momentum $\bp = \hbar\bk$. Needless to say, this is also what one obtains in the Minkowski formulation in the rest frame.

Thus we have $p = nE$ for a photon in the medium. Had we used the Abraham momentum density ${\bf E}\times{\bf H}$ instead, one would have obtained a photon momentum smaller by the factor $n^2$, i.e. $p = E/n$. This results in a corresponding smaller radiation pressure and seems to be ruled out by most experiments\cite{pressure}\cite{Brevik}\cite{Chiao}.

\section{Discussion}

Since the four-momentum of a particle with energy $E$ and momentum $\bp$ is $p^\mu = (nE,\bp)$ in our framework, the equivalent of the squared mass for a photon becomes $p^\mu p_\mu = 0$ when we make use of the previous result $E=p/n$. It is for this reason that excitations in solid-state physics with linear dispersion relations often are said to be mass-less.  On the other hand, in the original, covariant formalism where $p^\mu = (E,\bp)$, a free photon is seen to have a space-like  four-momentum in the Minkowski theory, i.e. like a tachyon. In this respect the situation is a bit better for the Abraham theory where the corresponding four-momentum is time-like, appropriate for a massive particle moving with a velocity less than the speed of light in vacuum. But with a massive photon, we would {\it a priori} expect problems with gauge invariance and more than the two ordinary  polarization degrees of freedom.

Zero-point oscillations of the electromagnetic field in a medium give rise to the energy $E_0 = \sum_\bk \hbar\om_\bk$.  Its effect can be measured by the attractive Casimir force between two parallel plates separated by such a material\cite{Casimir}. Now since $\om_\bk = |\bk|/n$ in a medium, we see that this zero-point energy will be smaller by a factor $n$ compared with the energy $\sum_\bk \hbar |\bk|$ in vacuum.  Recently, Brevik and Milton have used the Minkowski energy-momentum tensor to calculate the Casimir force in this case. After a more lengthy calculation, they find the same reduction of the force\cite{BM}. This is to be expected since the momentum density is the same in the Minkowski theory as in our effective theory. It is not clear what the Abraham theory will give for the Casimir force.

A related effect is the black-body radiation energy within a cavity filled with such a medium.  If the walls are kept at the temperature $T$ and the system is in thermodynamic equilibrium, the energy density is given by the usual expression
\be
       U/V = 2\int\!{d^3k\over(2\pi)^3}{\hbar\om_\bk\over e^{\beta\hbar\om_\bk} - 1}
\ee
with $\beta = 1/k_B T$ where $k_B$ is the Boltzmann constant. For the same reason as for the Casimir force, this now gives a result larger than the standard Stefan-Boltzmann formula by a factor $n^3$. It will be important to confirm this experimentally. The pressure in the radiation is one third of this energy density so the energy-momentum tensor remains traceless at finite temperature. 

In the book by Landau and Lifshitz the same result has been obtained from the use of fluctuating current correlators in the enclosing walls\cite{LL}. Afterwards it is simply stated that the same, increased radiation density can be obtained more directly by simply using $\om_\bk = |\bk|/n$. No comments are presented about the consistency with the Abraham theory which has previously been endorsed.

The effective theory seems to be very similar to the Minkowski theory in experimental consequences. It offers mainly a different theoretical outlook on electromagnetic phenomena in media and the resulting formulation is more compact and consistent. One can only hope that with this alternative theoretical framework the remaining experimental situations which still seem to prefer the Abraham formalism, can be resolved in a satisfactory way.

Dispersion have so far not been been discussed in the above. In the proposed effective theory this can be described by extending the free Lagrangian (\ref{Lag.3}) with dimension-6 interaction terms\cite{EFT}.  It is shown that such a description will be valid for photon energies below an upper energy cut-off in the range 5 - 10 eV. Similarly,  dimension-8 interactions in the effective Lagrangian can describe the different Kerr effects. At the quantum level these new interactions will have effects which can be calculated by standard perturbative methods of field theory.  They give rise to small corrections to the above lowest-order results for the Casimir force and black-body radiation density. In this way the effective theory relates many different classical and quantum optical phenomena into a unified description.

This work has been helped by discussions with I. Brevik, T. Curtright and L. Mezincescu.  Colleagues at the Department of Physics at UM are acknowledged for their hospitality, in particular T. von Hippel for also stressing the importance of experimental consequences of the effective theory.

\end{document}